\batchmode
\makeatletter
\def\input@path{{\string"/home/hari/Git/papers/Working/The Montecinos-Balsara ADER-FV Polynomial Basis - Convergence Properties & Extension to Non-Conservative Multidimensional Systems/\string"}}
\makeatother
\documentclass[twoside,english,final,5p,times,twocolumn]{elsarticle}
\usepackage[T1]{fontenc}
\usepackage[latin9]{inputenc}
\usepackage{amsmath}
\usepackage{amssymb}

\makeatletter
\journal{Computers and Fluids}

\usepackage{fancyhdr}
\makeatother

\usepackage{babel}
\begin{document}

\begin{frontmatter}{}

\title{The Montecinos-Balsara ADER-FV Polynomial Basis: Convergence Properties
\& Extension to Non-Conservative Multidimensional Systems}

\author{Haran Jackson}

\ead{hj305@cam.ac.uk}

\address{Cavendish Laboratory, JJ Thomson Ave, Cambridge, UK, CB3 0HE}
\begin{abstract}
Hyperbolic systems of PDEs can be solved to arbitrary orders of accuracy
by using the ADER Finite Volume method. These PDE systems may be non-conservative
and non-homogeneous, and contain stiff source terms. ADER-FV requires
a spatio-temporal polynomial reconstruction of the data in each spacetime
cell, at each time step. This reconstruction is obtained as the root
of a nonlinear system, resulting from the use of a Galerkin method.
It was proved in \citet{Jackson2017} that for traditional choices
of basis polynomials, the eigenvalues of certain matrices appearing
in these nonlinear systems are always 0, regardless of the number
of spatial dimensions of the PDEs or the chosen order of accuracy
of the ADER-FV method. This guarantees fast convergence to the Galerkin
root for certain classes of PDEs.

In \citet{Montecinos2017} a new, more efficient class of basis polynomials
for the one-dimensional ADER-FV method was presented. This new class
of basis polynomials, originally presented for conservative systems,
is extended to multidimensional, non-conservative systems here, and
the corresponding property regarding the eigenvalues of the Galerkin
matrices is proved.
\end{abstract}

\end{frontmatter}{}

\section{Background}

ADER-FV methods were first devised by Toro and collaborators (see
\citet{Millington2001}). \citet{Dumbser2008} obviated the need for
the cumbersome analytic work required by the Cauchy-Kowalesky procedure
by use of a Galerkin predictor. Although ADER-FV methods have been
very successful in solving a large variety of different hyperbolic
systems (e.g. see \citet{Dumbser2016a,Balsara2009,Hidalgo2011,Zanotti2016}),
they remain relatively computationally expensive.

\citet{Montecinos2017} have proposed a new, more efficient class
of basis polynomials. While the method was given for conservative,
one-dimensional systems in the original paper, it is extended here
to general non-conservative, multidimensional systems.

\section{Extension of the Montecinos-Balsara Formulation}

Take a non-homogeneous, non-conservative, hyperbolic system of the
form:

\begin{equation}
\frac{\partial\boldsymbol{Q}}{\partial t}+\boldsymbol{\nabla}\cdot\overrightarrow{\boldsymbol{F}}\left(\boldsymbol{Q}\right)+\overrightarrow{\boldsymbol{B}}\left(\boldsymbol{Q}\right)\cdot\nabla\boldsymbol{Q}=\boldsymbol{S}\left(\boldsymbol{Q}\right)\label{eq:system}
\end{equation}

where $\boldsymbol{Q}$ is the vector of conserved variables, $\overrightarrow{\boldsymbol{F}}=\left(\boldsymbol{F_{1},F_{2},F_{3}}\right)$
and $\overrightarrow{\boldsymbol{B}}=\left(B_{1},B_{2},B_{3}\right)$
are respectively the conservative nonlinear fluxes and matrices corresponding
to the purely non-conservative components of the system, and $\boldsymbol{S}\left(\boldsymbol{Q}\right)$
is the algebraic source vector.

Define spatial variables $x^{\left(1\right)},x^{\left(2\right)},x^{\left(3\right)}$.
Take the space-time cell $C=\left[x_{i_{1}}^{\left(1\right)},x_{i_{1}+1}^{\left(1\right)}\right]\times\left[x_{i_{2}}^{\left(2\right)},x_{i_{2}+1}^{\left(2\right)}\right]\times\left[x_{i_{3}}^{\left(3\right)},x_{i_{3}+1}^{\left(3\right)}\right]\times\left[t_{n},t_{n+1}\right]$.
Define the scaled spatial and temporal variables:

\begin{subequations}

\begin{equation}
\chi^{\left(k\right)}=\frac{x^{\left(k\right)}-x_{i_{k}}^{\left(k\right)}}{x_{i_{k}+1}^{\left(k\right)}-x_{i_{k}}^{\left(k\right)}}
\end{equation}

\begin{equation}
\tau=\frac{t-t_{n}}{t_{n+1}-t_{n}}
\end{equation}

\end{subequations}

Thus, $C$ becomes:

\begin{equation}
\left(\text{\ensuremath{\chi^{\left(1\right)}},\ensuremath{\chi^{\left(2\right)}},\ensuremath{\chi^{\left(3\right)}},\ensuremath{\tau}}\right)\in\left[0,1\right]^{4}
\end{equation}

By rescaling $\overrightarrow{\boldsymbol{F}},\overrightarrow{\boldsymbol{B}},\boldsymbol{S}$
by the appropriate constant factors, and defining $\tilde{\nabla}=\left(\partial_{\chi^{\left(1\right)}},\partial_{\chi^{\left(2\right)}},\partial_{\chi^{\left(3\right)}}\right)$,
within $C$ equation \eqref{eq:system} becomes:

\begin{equation}
\frac{\partial\boldsymbol{Q}}{\partial\tau}+\tilde{\boldsymbol{\nabla}}\cdot\overrightarrow{\boldsymbol{F}}\left(\boldsymbol{Q}\right)+\overrightarrow{\boldsymbol{B}}\left(\boldsymbol{Q}\right)\cdot\tilde{\nabla}\boldsymbol{Q}=\boldsymbol{S}\left(\boldsymbol{Q}\right)\label{eq:system rescaled}
\end{equation}

A basis $\left\{ \psi_{0},...,\psi_{N}\right\} $ of $P_{N}$ and
inner product $\left\langle \cdot,\cdot\right\rangle $ are now required
to produce a polynomial reconstruction of $\boldsymbol{Q}$ within
$C$. Traditionally, this basis has been chosen to be either nodal
($\psi_{i}\left(\alpha_{j}\right)=\delta_{ij}$ where $\left\{ \alpha_{0},\ldots,\alpha_{N}\right\} $
are a set of nodes, e.g. the Gauss-Legendre abscissae - see \citet{Dumbser2008}),
or modal (e.g. the Legendre polynomials - see \citet{Balsara2009}).

\citet{Montecinos2017} take the following approach. $\left\langle \cdot,\cdot\right\rangle $
is taken to be the usual integral product on $\left[0,1\right]$.
Supposing that $N=2n+1$ for some $n\in\mathbb{N}$, Gauss-Legendre
nodes $\left\{ \alpha_{0},\ldots,\alpha_{n}\right\} $ are taken.
The basis $\Psi=\left\{ \psi_{0},...,\psi_{N}\right\} \subset P_{N}$
is taken with the following properties for $i=0,\ldots,n$:

\begin{equation}
\begin{cases}
\psi_{i}\left(\alpha_{j}\right)=\delta_{ij} & \psi'_{i}\left(\alpha_{j}\right)=0\\
\psi_{n+1+i}\left(\alpha_{j}\right)=0 & \psi'_{n+1+i}\left(\alpha_{j}\right)=\delta_{ij}
\end{cases}
\end{equation}

Define the following subsets:

\begin{subequations}

\begin{align}
\Psi^{0} & =\left\{ \psi_{i}:0\leq i\leq n\right\} \\
\Psi^{1} & =\left\{ \psi_{i}:n+1\leq i\leq2n+1\right\} 
\end{align}

\end{subequations}

The WENO method (as used in \citet{Dumbser2013}) produces an order-$N$
polynomial reconstruction $w\left(\chi^{\left(1\right)},\chi^{\left(2\right)},\chi^{\left(3\right)}\right)$
of the data at time $t_{n}$ in $\left[x_{i_{1}}^{\left(1\right)},x_{i_{1}+1}^{\left(1\right)}\right]\times\left[x_{i_{2}}^{\left(2\right)},x_{i_{2}+1}^{\left(2\right)}\right]\times\left[x_{i_{3}}^{\left(3\right)},x_{i_{3}+1}^{\left(3\right)}\right]$.
It is used as initial data in the problem of finding the Galerkin
predictor. Taking representation $w=w_{abc}\psi_{a}\left(\chi^{\left(1\right)}\right)\psi_{b}\left(\chi^{\left(2\right)}\right)\psi_{c}\left(\chi^{\left(3\right)}\right)$
we have for $0\leq i,j,k\leq n$:

\begin{subequations}

\begin{align}
w_{ijk} & =w\left(\alpha_{i},\alpha_{j},\alpha_{k}\right)\\
w_{\left(n+i+1\right)jk} & =\partial_{\chi^{\left(1\right)}}w\left(\alpha_{i},\alpha_{j},\alpha_{k}\right)\\
w_{i\left(n+j+1\right)k} & =\partial_{\chi^{\left(2\right)}}w\left(\alpha_{i},\alpha_{j},\alpha_{k}\right)\\
w_{ij\left(n+k+1\right)} & =\partial_{\chi^{\left(3\right)}}w\left(\alpha_{i},\alpha_{j},\alpha_{k}\right)
\end{align}

\end{subequations}

Take the following temporal nodes, where $\tau_{1},\ldots,\tau_{N}$
are the usual Legendre-Gauss nodes on $\left[0,1\right]$ and $\tau_{0}=0$
or $\tau_{0}=1$ if we are performing a Continuous Galerkin / Discontinuous
Galerkin reconstruction, respectively:

\begin{equation}
\left\{ \tau_{0},\dots,\tau_{N}\right\} 
\end{equation}

Define $\Phi=\left\{ \phi_{0},...,\phi_{N}\right\} \subset P_{N}$
to be the set of Lagrange interpolating polynomials on the temporal
nodes. We now define the spatio-temporal polynomial basis $\Theta=\Phi\otimes\Psi\otimes\Psi\otimes\Psi=\left\{ \theta_{\beta}\right\} $
for $0\leq\beta\leq\left(N+1\right)^{4}-1$. Define subsets $\Theta^{\iota\xi\kappa}=\Phi\otimes\Psi^{\iota}\otimes\Psi^{\xi}\otimes\Psi^{\kappa}=\left\{ \theta_{\mu}^{\iota\xi\kappa}\right\} $
where $\iota,\xi,\kappa\in\left\{ 0,1\right\} $ for $0\leq\mu\leq\left(N+1\right)\left(n+1\right)^{3}-1$.

Denoting the Galerkin predictor by $\boldsymbol{q}$, take the following
set of approximations:

\begin{subequations}

\begin{align}
\boldsymbol{Q} & \approx\theta_{\beta}\boldsymbol{q}_{\beta}=\theta_{\mu}^{\iota\xi\kappa}\boldsymbol{q}_{\mu}^{\iota\xi\kappa}\\
\overrightarrow{\boldsymbol{F}}\left(\boldsymbol{Q}\right) & \approx\theta_{\beta}\overrightarrow{\boldsymbol{F}}_{\beta}=\theta_{\mu}^{\iota\xi\kappa}\overrightarrow{\boldsymbol{F}}_{\mu}^{\iota\xi\kappa}\label{eq:F representation}\\
\overrightarrow{\boldsymbol{B}}\left(\boldsymbol{Q}\right)\cdot\tilde{\nabla}\boldsymbol{Q} & \approx\theta_{\beta}\boldsymbol{B}_{\beta}=\theta_{\mu}^{\iota\xi\kappa}\boldsymbol{B}_{\mu}^{\iota\xi\kappa}\label{eq:B representation}\\
\boldsymbol{S}\left(\boldsymbol{Q}\right) & \approx\theta_{\beta}\boldsymbol{S}_{\beta}=\theta_{\mu}^{\iota\xi\kappa}\boldsymbol{S}_{\mu}^{\iota\xi\kappa}\label{eq:S representation}
\end{align}

\end{subequations}

for some coefficients $\boldsymbol{q}_{\beta},\overrightarrow{\boldsymbol{F}}_{\beta},\boldsymbol{B}_{\beta},\boldsymbol{S}_{\beta}$.
The \textit{nodal basis representation} is used for the coefficients
of $\Theta^{000}$:

\begin{subequations}

\begin{align}
\overrightarrow{\boldsymbol{F}}_{\mu}^{000} & =\overrightarrow{\boldsymbol{F}}\left(\boldsymbol{q}_{\mu}^{000}\right)\\
\boldsymbol{B}_{\mu}^{000} & =B_{1}\left(\boldsymbol{q}_{\mu}^{000}\right)\boldsymbol{q}_{\mu}^{100}+B_{2}\left(\boldsymbol{q}_{\mu}^{000}\right)\boldsymbol{q}_{\mu}^{010}+B_{3}\left(\boldsymbol{q}_{\mu}^{000}\right)\boldsymbol{q}_{\mu}^{001}\\
\boldsymbol{S}_{\mu}^{000} & =\boldsymbol{S}\left(\boldsymbol{q}_{\mu}^{000}\right)
\end{align}

\end{subequations}

In general, we have:

\begin{subequations}

\begin{align}
\overrightarrow{\boldsymbol{F}}_{\mu}^{\iota\xi\kappa} & =\partial_{\chi}^{\iota}\partial_{\upsilon}^{\xi}\partial_{\zeta}^{\kappa}\left(\overrightarrow{\boldsymbol{F}}\left(\boldsymbol{Q}\right)\right)\\
\boldsymbol{B}_{\mu}^{\iota\xi\kappa} & =\partial_{\chi}^{\iota}\partial_{\upsilon}^{\xi}\partial_{\zeta}^{\kappa}\left(\overrightarrow{\boldsymbol{B}}\left(\boldsymbol{Q}\right)\cdot\tilde{\nabla}\boldsymbol{Q}\right)\\
\boldsymbol{S}_{\mu}^{\iota\xi\kappa} & =\partial_{\chi}^{\iota}\partial_{\upsilon}^{\xi}\partial_{\zeta}^{\kappa}\left(\boldsymbol{S}\left(\boldsymbol{Q}\right)\right)\label{eq:S coeffs}
\end{align}

\end{subequations}

where the right-hand-side is evaluated at the nodal point corresponding
to $\mu$. The full expressions are omitted here for brevity's sake,
but note that for a one-dimensional system:

\begin{subequations}

\begin{align}
\boldsymbol{F}_{1}{}_{\mu}^{100} & =\frac{\partial\boldsymbol{F}\left(\boldsymbol{q}_{\mu}^{000}\right)}{\partial\boldsymbol{Q}}\cdot\boldsymbol{q}_{\mu}^{100}\\
\boldsymbol{B}_{\mu}^{100} & =\left(\frac{\partial B_{1}\left(\boldsymbol{q}_{\mu}^{000}\right)}{\partial\boldsymbol{Q}}\cdot\boldsymbol{q}_{\mu}^{100}\right)\cdot\boldsymbol{q}_{\mu}^{100}\\
 & +B_{1}\left(\boldsymbol{q}_{\mu}^{000}\right)\cdot\left(\frac{\partial^{2}\theta_{\kappa}^{000}\left(\chi_{\mu},\tau_{\mu}\right)}{\partial\chi^{2}}\boldsymbol{q}_{\mu}^{000}+\frac{\partial^{2}\theta_{\kappa}^{100}\left(\chi_{\mu},\tau_{\mu}\right)}{\partial\chi^{2}}\boldsymbol{q}_{\mu}^{100}\right)\nonumber \\
\boldsymbol{S}_{\mu}^{100} & =\frac{\partial\boldsymbol{S}\left(\boldsymbol{q}_{\mu}^{000}\right)}{\partial\boldsymbol{Q}}\cdot\boldsymbol{q}_{\mu}^{100}
\end{align}

\end{subequations}

where $\chi_{\mu},\tau_{\mu}$ are the spatial and temporal coordinates
where $\theta_{\mu}^{100}=0$ and $\partial_{\chi}\theta_{\mu}^{100}=1$.
Note that $\frac{\partial B_{1}}{\partial\boldsymbol{Q}}$ is a rank
3 tensor.

Consider functions $f,g$ of the following form:

\begin{subequations}

\begin{align}
f\left(\tau,\chi^{\left(1\right)},\chi^{\left(2\right)},\chi^{\left(3\right)}\right) & =f_{\tau}\left(\tau\right)f_{1}\left(\chi^{\left(1\right)}\right)f_{2}\left(\chi^{\left(2\right)}\right)f_{3}\left(\chi^{\left(3\right)}\right)\\
g\left(\tau,\chi^{\left(1\right)},\chi^{\left(2\right)},\chi^{\left(3\right)}\right) & =g_{\tau}\left(\tau\right)g_{1}\left(\chi^{\left(1\right)}\right)g_{2}\left(\chi^{\left(2\right)}\right)g_{3}\left(\chi^{\left(3\right)}\right)
\end{align}

\end{subequations}

Define the following integral operators:

\begin{subequations}

\begin{align}
\left[f,g\right]^{t} & =f_{\tau}\left(t\right)g_{\tau}\left(t\right)\left\langle f_{1},g_{1}\right\rangle \left\langle f_{2},g_{2}\right\rangle \left\langle f_{3},g_{3}\right\rangle \\
\left\{ f,g\right\}  & =\left\langle f_{\tau},g_{\tau}\right\rangle \left\langle f_{1},g_{1}\right\rangle \left\langle f_{2},g_{2}\right\rangle \left\langle f_{3},g_{3}\right\rangle 
\end{align}

\end{subequations}

Multiplying \eqref{eq:F representation} by test function $\theta_{\alpha}$,
using the polynomial approximations for $\boldsymbol{Q},\overrightarrow{\boldsymbol{F}},\overrightarrow{\boldsymbol{B}},\boldsymbol{S}$,
and integrating over space and time gives:

\begin{equation}
\left\{ \theta_{\alpha},\frac{\partial\theta_{\beta}}{\partial\tau}\right\} \boldsymbol{q}_{\beta}=\left\{ \theta_{\alpha},\theta_{\beta}\right\} \left(\boldsymbol{S}_{\beta}-\boldsymbol{B}_{\beta}\right)-\left\{ \theta_{\alpha},\frac{\partial\theta_{\beta}}{\partial\chi^{\left(k\right)}}\right\} \boldsymbol{F}_{k}{}_{\beta}\label{eq:conditions}
\end{equation}

\subsection{The Discontinuous Galerkin Method}

This method of computing the Galerkin predictor allows solutions to
be discontinuous at temporal cell boundaries, and is also suitable
for stiff source terms. Integrating \eqref{eq:conditions} by parts
in time gives:

\begin{align}
\left(\left[\theta_{\alpha},\theta_{\beta}\right]^{1}-\left\{ \frac{\partial\theta_{\alpha}}{\partial\tau},\theta_{\beta}\right\} \right)\boldsymbol{q}_{\beta} & =\left[\theta_{\alpha},\boldsymbol{w}\right]^{0}+\left\{ \theta_{\alpha},\theta_{\beta}\right\} \left(\boldsymbol{S}_{\beta}-\boldsymbol{B}_{\beta}\right)\\
 & -\left\{ \theta_{\alpha},\frac{\partial\theta_{\beta}}{\partial\chi^{\left(k\right)}}\right\} \boldsymbol{F}_{k}{}_{\beta}\nonumber 
\end{align}

where $\boldsymbol{w}$ is the reconstruction obtained at the start
of the time step with the WENO method. Take the following ordering:

\begin{equation}
\theta_{\left(N+1\right)^{3}h+\left(N+1\right)^{2}i+\left(N+1\right)j+k}\left(\tau,\chi,\upsilon,\zeta\right)=\phi_{h}\left(\tau\right)\psi_{i}\left(\chi\right)\psi_{j}\left(\upsilon\right)\psi_{k}\left(\zeta\right)
\end{equation}

where $0\leq h,i,j,k\leq N$. Thus, define the following:

\begin{subequations}

\begin{align}
U_{\alpha\beta} & =\left[\theta_{\alpha},\theta_{\beta}\right]^{1}-\left\{ \frac{\partial\theta_{\alpha}}{\partial\tau},\theta_{\beta}\right\} =\left(R^{1}-M^{\tau,1}\right)\otimes\left(M^{\chi}\right)^{3}\\
V_{\alpha\beta}^{k} & =\left\{ \theta_{\alpha},\frac{\partial\theta_{\beta}}{\partial\chi^{\left(k\right)}}\right\} =M^{\tau}\otimes\left(M^{\chi}\right)^{k-1}\otimes M^{\chi,1}\otimes\left(M^{\chi}\right)^{3-k}\\
\boldsymbol{W}_{\alpha} & =\left[\theta_{\alpha},\Psi_{\gamma}\right]^{0}\boldsymbol{w}_{\gamma}=R^{0}\otimes\left(M^{\chi}\right)^{3}\\
Z_{\alpha\beta} & =\left\{ \theta_{\alpha},\theta_{\beta}\right\} =M^{\tau}\otimes\left(M^{\chi}\right)^{3}
\end{align}

\end{subequations}

where $\left\{ \Psi_{\gamma}\right\} =\Psi\otimes\Psi\otimes\Psi$
and:

\begin{equation}
\begin{cases}
M_{ij}^{\tau}=\left\langle \phi_{i},\phi_{j}\right\rangle  & M_{ij}^{\tau,1}=\left\langle \phi_{i}^{'},\phi_{j}\right\rangle \\
M_{ij}^{\chi}=\left\langle \psi_{i},\psi_{j}\right\rangle  & M_{ij}^{\chi,1}=\left\langle \psi_{i},\psi_{j}^{'}\right\rangle \\
R_{ij}^{1}=\phi_{i}\left(1\right)\phi_{j}\left(1\right) & \boldsymbol{R_{i}^{0}}=\phi_{i}\left(0\right)
\end{cases}\label{eq:inner product matrices}
\end{equation}

Thus:

\begin{equation}
U_{\alpha\beta}\boldsymbol{q}_{\beta}=\boldsymbol{W}_{\alpha}+Z_{\alpha\beta}\left(\boldsymbol{S}_{\beta}-\boldsymbol{B}_{\beta}\right)-V_{\alpha\beta}^{\left(k\right)}\boldsymbol{F}_{k}{}_{\beta}
\end{equation}

Take the definitions:

\begin{equation}
\begin{cases}
D & =\left(M^{\chi}\right)^{-1}M^{\chi,1}\\
E & =\left(R^{1}-M^{\tau,1}\right)
\end{cases}\label{eq:compound matrices}
\end{equation}

Noting that $E\boldsymbol{1}=\boldsymbol{R^{0}}$, we have, by inversion
of $U$:

\begin{align}
\boldsymbol{q} & =\left(\boldsymbol{1}\otimes I^{3}\right)\boldsymbol{w}+\left(E^{-1}M^{\tau}\otimes I^{3}\right)\left(\boldsymbol{S}-\boldsymbol{B}\right)\\
 & -\left(E^{-1}M^{\tau}\otimes I^{k-1}\otimes D\otimes I^{3-k}\right)\boldsymbol{F}_{k}\nonumber 
\end{align}

Thus, we have:

\begin{align}
\boldsymbol{q}_{hijk} & =\boldsymbol{w}_{ijk}+\left(E^{-1}M^{\tau}\right)_{hm}\left(\boldsymbol{S}_{mijk}-\boldsymbol{B}_{mijk}\right)\label{eq:q system}\\
 & -\left(E^{-1}M^{\tau}\right)_{hm}\left(D_{in}\left(\boldsymbol{F}_{1}\right)_{mnjk}+D_{jn}\left(\boldsymbol{F}_{2}\right)_{mink}+D_{kn}\left(\boldsymbol{F}_{3}\right)_{mijn}\right)\nonumber 
\end{align}

Note then that $\boldsymbol{q}^{\iota\xi\kappa}$ is a function of
$\boldsymbol{S}^{\iota\xi\kappa},\boldsymbol{B}^{\iota\xi\kappa},\overrightarrow{\boldsymbol{F}}$:

\begin{equation}
\boldsymbol{q}^{\iota\xi\kappa}=\mathcal{F}\left(\boldsymbol{S}^{\iota\xi\kappa}\right)+\mathcal{F}\left(\boldsymbol{B}^{\iota\xi\kappa}\right)+\mathcal{G_{\iota\xi\kappa}}\left(\overrightarrow{\boldsymbol{F}}^{000},\ldots,\overrightarrow{\boldsymbol{F}}^{111}\right)
\end{equation}

where $\mathcal{F},\mathcal{G_{\iota\xi\kappa}}$ are linear functions.
Note in turn that, by \eqref{eq:S coeffs}:

\begin{equation}
\boldsymbol{S}^{\iota\xi\kappa}=\mathcal{H}\left(\bigcup_{\begin{array}{c}
\left(0,0,0\right)\leq\left(a,b,c\right)\leq\left(\iota,\xi,\kappa\right)\end{array}}\boldsymbol{q}^{abc}\right)
\end{equation}

where $\mathcal{H}$ is a nonlinear function.

In the case of stiff source terms, the following Picard iteration
procedure can be used to solve \eqref{eq:q system}, as adapted from
\citet{Montecinos2017}:

\begin{align}
\left(\boldsymbol{q}^{\iota\xi\kappa}\right)_{m+1} & =\mathcal{F}\left(\mathcal{H}\left(\left(\boldsymbol{q}^{\iota\xi\kappa}\right)_{m+1}\cup\bigcup_{\begin{array}{c}
\left(0,0,0\right)\leq\left(a,b,c\right)\leq\left(\iota,\xi,\kappa\right)\\
\left(a,b,c\right)\neq\left(\iota,\xi,\kappa\right)
\end{array}}\left(\boldsymbol{q}^{abc}\right)_{m}\right)\right)\\
 & +\mathcal{F}\left(\left(\boldsymbol{B}^{\iota\xi\kappa}\right)_{m}\right)+\mathcal{G_{\iota\xi\kappa}}\left(\left(\overrightarrow{\boldsymbol{F}}^{000}\right)_{m},\ldots,\left(\overrightarrow{\boldsymbol{F}}^{111}\right)_{m}\right)\nonumber 
\end{align}

\subsection{The Continuous Galerkin Method}

This method of computing the Galerkin predictor is not suitable for
stiff source terms, but is less computationally expensive and ensures
continuity across temporal cell boundaries. The first $N+1$ elements
of $\boldsymbol{q}$ are fixed by imposing the following condition:

\begin{equation}
\boldsymbol{q}\left(\chi,0\right)=\boldsymbol{w}\left(\chi\right)
\end{equation}

For $\boldsymbol{v}\in\mathbb{R}^{\left(N+1\right)^{2}}$ and $X\in M_{\left(N+1\right)^{2},\left(N+1\right)^{2}}\left(\mathbb{R}\right)$,
let $\boldsymbol{v}=\left(\boldsymbol{v^{0}},\boldsymbol{v^{1}}\right)$
and $X=\left(\begin{array}{cc}
X^{00} & X^{01}\\
X^{10} & X^{11}
\end{array}\right)$ where $\boldsymbol{v^{0}},X^{00}$ are the components relating solely
to the first $N+1$ components of $\boldsymbol{v}$. We only need
to find the latter components of $\boldsymbol{q}$, and thus, from
\eqref{eq:conditions}, we have:

\begin{align}
\left\{ \theta_{\alpha},\frac{\partial\theta_{\beta}}{\partial\tau}\right\} ^{11}\boldsymbol{q}_{\beta}^{1} & =\left\{ \theta_{\alpha},\theta_{\beta}\right\} ^{11}\left(\boldsymbol{S}_{\beta}^{1}-\boldsymbol{B}_{\beta}^{1}\right)-\left\{ \theta_{\alpha},\frac{\partial\theta_{\beta}}{\partial\chi^{\left(k\right)}}\right\} ^{11}\boldsymbol{F}_{k}{}_{\beta}^{1}\\
 & +\left\{ \theta_{\alpha},\theta_{\beta}\right\} ^{10}\left(\boldsymbol{S}_{\beta}^{0}-\boldsymbol{B}_{\beta}^{0}\right)-\left\{ \theta_{\alpha},\frac{\partial\theta_{\beta}}{\partial\chi^{\left(k\right)}}\right\} ^{10}\boldsymbol{F}_{k}{}_{\beta}^{0}\nonumber 
\end{align}
Define the following:

\begin{subequations}

\begin{align}
U_{\alpha\beta} & =\left\{ \theta_{\alpha},\frac{\partial\theta_{\beta}}{\partial\tau}\right\} ^{11}\\
V_{\alpha\beta}^{k} & =\left\{ \theta_{\alpha},\frac{\partial\theta_{\beta}}{\partial\chi^{\left(k\right)}}\right\} ^{11}\\
\boldsymbol{W}_{\alpha} & =\left\{ \theta_{\alpha},\theta_{\beta}\right\} ^{10}\left(\boldsymbol{S}_{\beta}-\boldsymbol{B}_{\beta}\right)^{0}-\left\{ \theta_{\alpha},\frac{\partial\theta_{\beta}}{\partial\chi^{\left(k\right)}}\right\} ^{10}\boldsymbol{F}_{k}{}_{\beta}^{0}\\
Z_{\alpha\beta} & =\left\{ \theta_{\alpha},\theta_{\beta}\right\} ^{11}
\end{align}

\end{subequations}

Thus:

\begin{equation}
U_{\alpha\beta}\boldsymbol{q}_{\beta}^{1}=\boldsymbol{W}_{\alpha}+Z_{\alpha\beta}\left(\boldsymbol{S}_{\beta}^{1}-\boldsymbol{B}_{\beta}^{1}\right)-V_{\alpha\beta}^{k}\boldsymbol{F}_{k}{}_{\beta}^{1}
\end{equation}

Note that, as with the discontinuous Galerkin method, $\boldsymbol{W}$
has no dependence on the degrees of freedom in $\boldsymbol{q}$.
As the source terms are not stiff, the following iteration is used:

\begin{equation}
U_{\alpha\beta}\left(\boldsymbol{q}_{\beta}^{1}\right)_{m+1}=\boldsymbol{W}_{\alpha}+Z_{\alpha\beta}\left(\left(\boldsymbol{S}_{\beta}^{1}\right)_{m}-\left(\boldsymbol{B}_{\beta}^{1}\right)_{m}\right)-V_{\alpha\beta}^{k}\left(\boldsymbol{F}_{k}{}_{\beta}^{1}\right)_{m}
\end{equation}

\section{Convergence Properties}

In \citet{Jackson2017} it was proved that for traditional choices
of polynomial bases, the eigenvalues of $U^{-1}V^{i}$ are all 0 for
any $N\in\mathbb{N}$, for $i=1,2,3$. This implies that in the conservative,
homogeneous case ($\overrightarrow{\boldsymbol{B}}=\boldsymbol{S}=\boldsymbol{0}$),
owing to the Banach Fixed Point Theorem, existence and uniqueness
of a solution are established, and convergence to this solution is
guaranteed. As noted in \citet{Dumbser2009a}, in the linear case
it is implied that the iterative procedure converges after at most
$N+1$ iterations. A proof of this result for the Montecinos-Balsara
polynomial basis class is now provided here. For the theory in linear
algebra required for this section, please consult a standard textbook
on the subject, such as \citet{Nering1970}.

Take the definitions \eqref{eq:inner product matrices}, \eqref{eq:compound matrices}.
Consider that:

\begin{equation}
U^{-1}V^{k}=E^{-1}M^{\tau}\otimes I^{k-1}\otimes D\otimes I^{3-k}
\end{equation}

Therefore:

\begin{equation}
\left(U^{-1}V^{k}\right)^{m}=\left(E^{-1}M^{\tau}\right)^{m}\otimes\left(I^{k-1}\right)^{m}\otimes D^{m}\otimes\left(I^{3-k}\right)^{m}
\end{equation}

A matrix $X$ is nilpotent ($X^{k}=0$ for some $k\in\mathbb{N}$)
if and only if all its eigenvalues are 0. Note that $U^{-1}V^{k}$
is nilpotent if $D^{m}=0$ for some $m\in\mathbb{N}$.

Note that if $p\in P_{N}$ then $p=a_{j}\psi_{j}$ for some unique
coefficient vector $\boldsymbol{a}$. Thus, taking inner products
with $\psi_{i}$, we have $\left\langle \psi_{i},\psi_{j}\right\rangle a_{j}=\left\langle \psi_{i},p\right\rangle $
for $i=0,...,N$. This produces the following result:

\begin{equation}
p=a_{j}\psi_{j}\Leftrightarrow\boldsymbol{a}=\left(M^{\chi}\right)^{-1}\boldsymbol{x},\;x_{i}=\left\langle \psi_{i},p\right\rangle \label{eq:gramCoeffs}
\end{equation}
 Taking $\boldsymbol{a}\in\mathbb{R}^{N+1}$, define:

\begin{equation}
p=a_{0}\psi_{0}+\ldots+a_{N}\psi_{N}\in P_{N}
\end{equation}

Note that:

\begin{equation}
\left(M^{\chi,1}\boldsymbol{a}\right)_{i}=\left\langle \psi_{i},\psi_{0}^{'}\right\rangle a_{0}+\ldots+\left\langle \psi_{i},\psi_{N}^{'}\right\rangle a_{N}=\left\langle \psi_{i},p^{'}\right\rangle 
\end{equation}

Thus, by \eqref{eq:gramCoeffs}:

\begin{equation}
\left(\left(M^{\chi}\right)^{-1}M^{\chi,1}\boldsymbol{a}\right)_{i}\psi_{i}=\left(D\boldsymbol{a}\right)_{i}\psi_{i}=p'
\end{equation}

By induction:

\begin{equation}
\left(D^{m}\boldsymbol{a}\right)_{i}\psi_{i}=p^{\left(m\right)}
\end{equation}

for any $m\in\mathbb{N}$. As $p\in P_{N}$, $D^{N+1}\boldsymbol{a}=\boldsymbol{0}$.
As $\boldsymbol{a}$ was chosen arbitrarily, $D^{N+1}=0$. No specific
choice has been made for $N\in\mathbb{N}$ and thus the result holds
in general.

\section{References}

\bibliographystyle{elsarticle-harv}
\addcontentsline{toc}{section}{\refname}\bibliography{0_home_hari_Git_papers_Working_The_Montecinos-B___-Conservative_Multidimensional_Systems_refs}

\section{Acknowledgments}

I acknowledge financial support from the EPSRC Centre for Doctoral
Training in Computational Methods for Materials Science under grant
EP/L015552/1.
\end{document}